\documentclass[aip,jcp,preprint,superscriptaddress]{revtex4-1}

\usepackage{amsmath}    
\usepackage{graphicx}   
\usepackage{color}      
\usepackage{subfigure}  
\usepackage{hyperref}   
\usepackage[latin1]{inputenc}

\begin{document}

\title{Absence of photoemission from the Fermi level in potassium intercalated picene and coronene films: structure, polaron or correlation physics?}
\author{Benjamin Mahns}
\author{Friedrich Roth}
\author{Martin Knupfer}
\affiliation{IFW Dresden, P.O. Box 270116, D-01171 Dresden, Germany}
\date{\today}

\begin{abstract}
The electronic structure of potassium intercalated picene and coronene films has been studied using photoemission spectroscopy. Picene has additionally been intercalated using sodium. Upon alkali metal addition core level as well as valence band photoemission data signal a filling of previously unoccupied states of the two molecular materials due to charge transfer from potassium. In contrast to the observation of superconductivity in K$_x$picene and K$_x$coronene  ($x \sim 3$), none
of the films studied shows emission from the Fermi level, i.\,e. we find no indication for a metallic ground state. Several reasons for this
observation are discussed.
\end{abstract}

\maketitle

\section{Introduction}

Superconductivity always has attracted a large number of researchers since this phenomenon harbors challenges and prospects both under
fundamental and applied points of view. Very recently, it has been discovered that some molecular crystal consisting of polycyclic aromatic
hydrocarbons demonstrate superconductivity upon alkali metal addition. Furthermore, these compounds are characterized by rather high transition
temperatures  into the superconducting state, for instance K$_3$picene ($T_c$=18\,K)\cite{Mitsuhashi2010} and K$_3$coronene
($T_c$=15\,K)\cite{Kubozono2011}. Most recently, superconductivity with a transition temperature of 33\,K has been reported for
K$_3$dibenzopentacene .\cite{Mianqi2011} These doped aromatic hydrocarbons thus represent a class of organic superconductors with transition temperatures
only slightly below that of the famous alkali metal doped fullerenes with $T_c$'s up to 40\,K
\cite{Hebard1991,Tanigaki1991,Palstra1995,Gunnarsson2004,Gunnarsson1997}.

\par

The development of a thorough understanding of the normal as well as superconducting state properties requires an investigation of the physical
properties of the corresponding molecular crystals in the undoped and doped state. In this contribution we present the investigation of the
occupied electronic states of two of these systems, picene and coronene, which have been grown as thin films on SiO$_2$ substrates and
subsequently doped by potassium or sodium addition.

\par

\begin{figure}[h]
\centering
\includegraphics[width=0.49\linewidth]{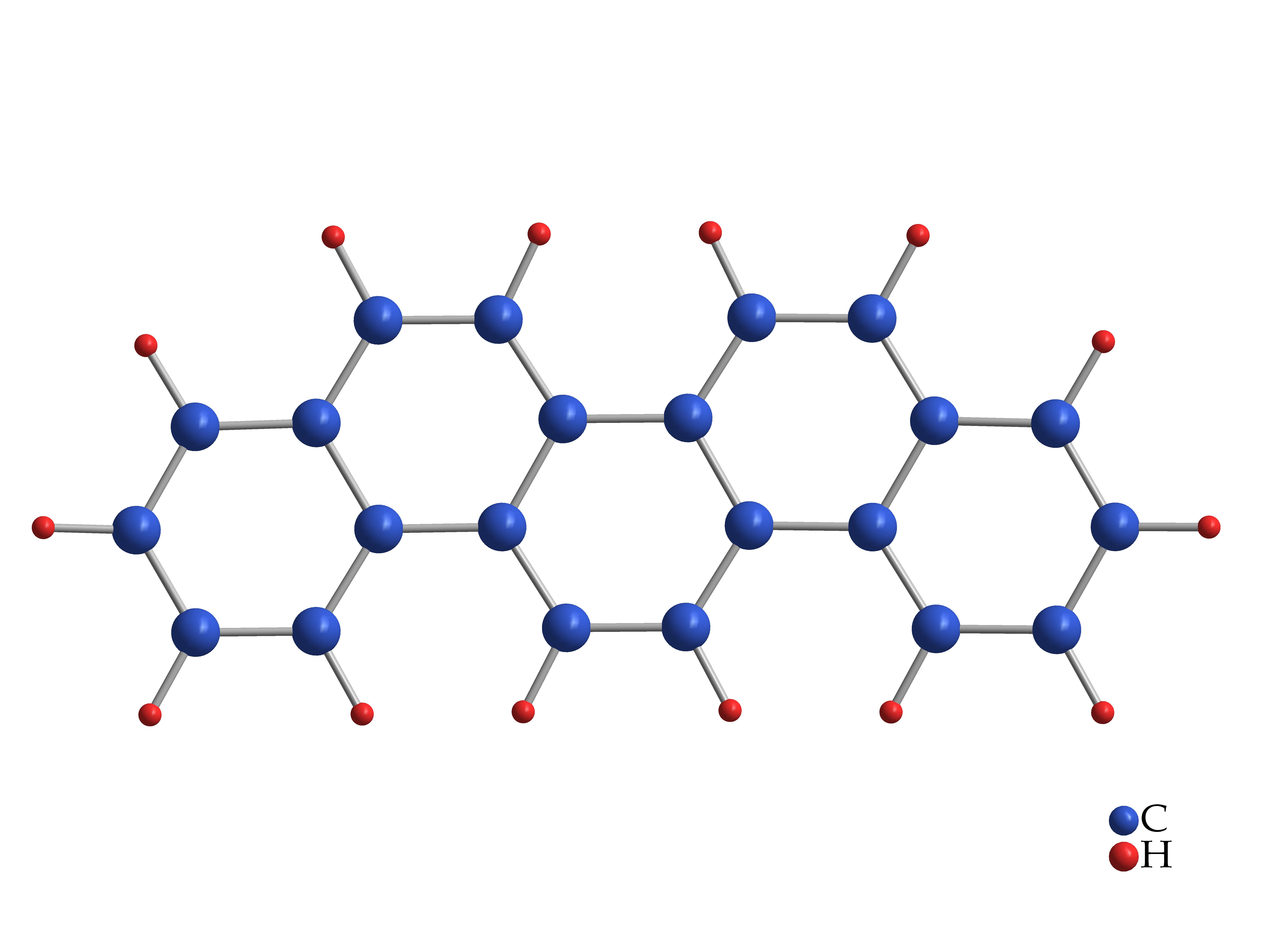}
\includegraphics[width=0.49\linewidth]{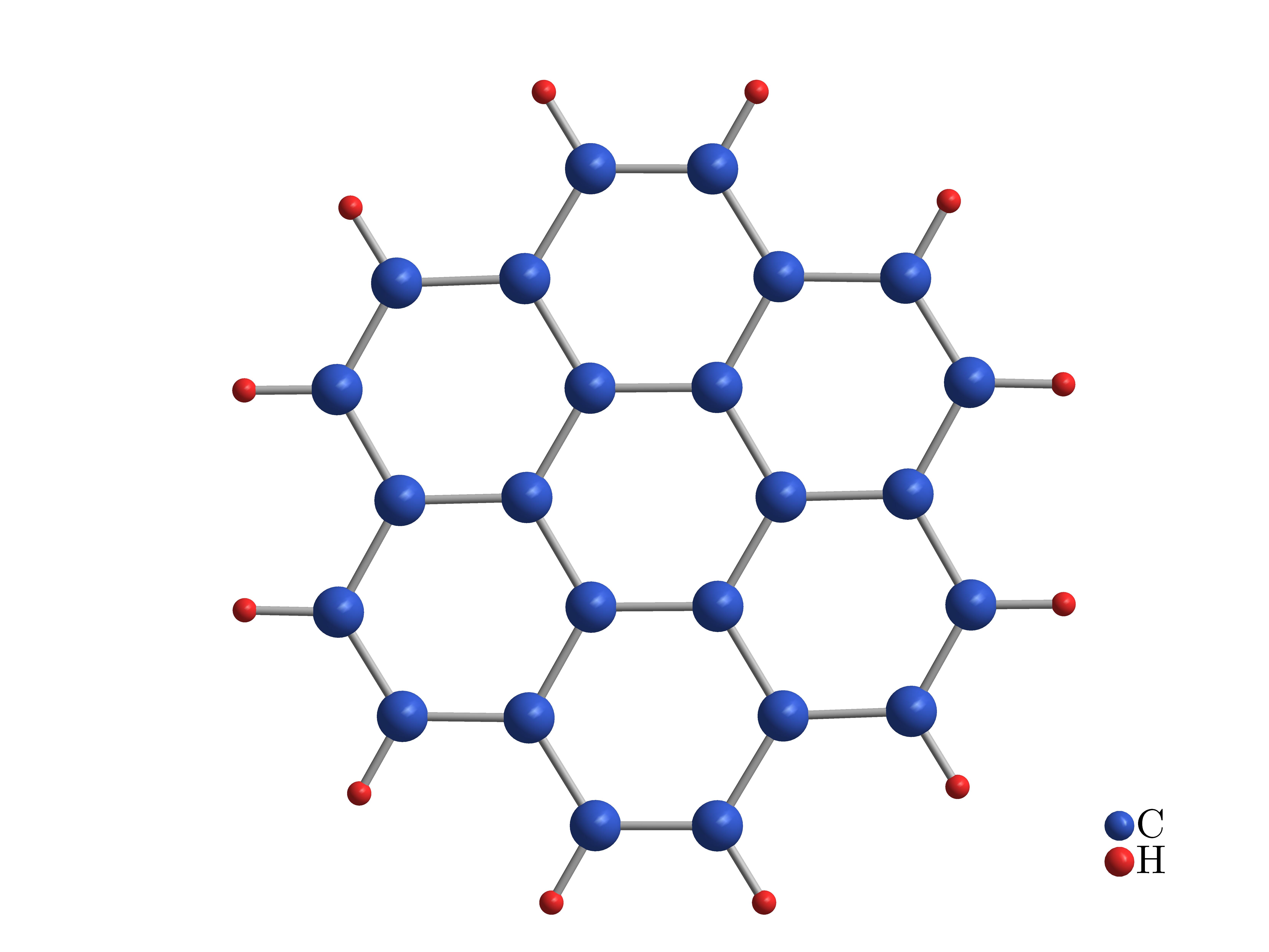}
\caption{Schematic representation of the molecular structure of
picene (left panel) and coronene (right panel).} \label{figstruc}
\end{figure}

Picene and coronene are molecules that consist of five and six benzene rings arranged in a zig-zack and ring-like manner, respectively. Fig.\,\ref{figstruc}
depicts a schematic representation of the two molecular strutures. Both materials adopt a monoclinic crystal structure in the condensed phase,
the lattice parameters are $a$ = 8.480\,\AA, $b$ = 6.154\,\AA, $c$ = 13.515 \,\AA, and $\beta$ = 90.46$^\circ$ for picene, and $a$ =
16.094\,\AA, $b$ = 4.690\,\AA, $c$ = 10.049 \,\AA, and $\beta$ = 110.79$^\circ$, for coronene.\cite{Echigo2007,De1985} In both cases the unit
cell contains two inequivalent molecules and the molecules arrange in a herringbone manner. The evolution of the crystal structure upon doping
is unknown in detail, experimental data that allow a detailed structure refinement have not been published yet. Calculations of Kubozono
\textit{et al.} indicate the intercalation of K atoms into picene within the ab-plane.\cite{Kubozono2011} This result is also supported by
first-principles structure optimization from Kosugi\textit{et al.} where the lowest total energy was found for dopants in the intralayer
region.\cite{Kosugi2011a,Andres2011} This behavior is different to K$_x$pentacene where experimental data indicate an intercalation of K into
the space between the $ab$-layers.\cite{Ito2004} A reduction of the unit cell volume V due to distortion of picene or a change of the molecule
orientation in the herringbone arrangement has also been found theoretically.\cite{Kosugi2011a} For K$_3$coronene a first DFT calculation wit K
atoms within the $ab$-plane results in a significant change of the heringbone structure and leads also to strong deformations of the molecule
itself. \cite{Kosugi2011}

\par

The electronic properties of the novel hydrocarbon superconductors has been addressed by a number of theory articles recently. Several aspects
like electron-phonon coupling, magnetic properties, electronic structure and correlation of the electron system have been
studied.\cite{Subedi2011,Kim2011,Casula2011,Kato2011,Kosugi2011,Roth2010,Cudazzo2011} Experimentally, there are only a few studies that deal with the
electronic structure of K$_x$picene. A first photoemission study has shown that upon potassium addition to picene films, a new structure in the
former energy gap is formed, which has been interpreted in terms of filled molecular states. Using electron-energy loss spectroscopy we have
recently addressed the electronic structure of undoped and potassium doped picene. These studies revealed that pristine picene is characterized
by four very close lying conduction bands and several excitonic features in the electronic excitation spectrum. Potassium addition leads to a
filling of the close lying conduction bands and causes the appearance of a new excitation features in the former band gap which can be
associated with the charge carrier plasmon in K$_3$picene. Equivalent changes have also been reported for potassium doped
coronene.\cite{Okazaki2010,Roth2010,Roth2011,Roth2011a,Roth2012}

\par

Surprisingly, our photoemission studies show that there is charge transfer from the added potassium atoms to the two molecules but, in contrast
to the report of superconductivity, our valence band data do not show emission from the Fermi level. This is discussed in the framework of
structural, phonon related and correlation effects.

\section{Experimental}

The X-ray (XPS) and ultra-violet (UPS) photoemission spectroscopy experiments have been carried out using a SPECS surface analysis system
containing a sample preparation and measuring chamber, each witch a base pressure lower than $4\times10^{-10}$\,mbar. The system is equipped
with an electron-energy analyzer PHOIBOS-150 (SPECS) and two light sources, respectively. A monochromatized Al K$_{\alpha}$ source provides
photons with an energy of 1486.6\,eV for XPS. Photons with an energy of 21.21\,eV from a He discharge lamp were used to perform valence band
measurements.

\par

The UPS measurements were done by applying a sample bias of -5\,eV
to obtain the correct secondary electron cutoff. The recorded
spectra were corrected for the contribution of He satellite
radiation. For XPS the energy scale was calibrated to reproduce the binding energy of
Au 4$f_{7/2}$ (84.0\,eV). The total energy resolution of the spectrometer was
about 0.35\,eV (XPS) and 0.06\,eV (UPS).

\par

\begin{figure}[h]
\centering
\includegraphics[width=0.8\linewidth]{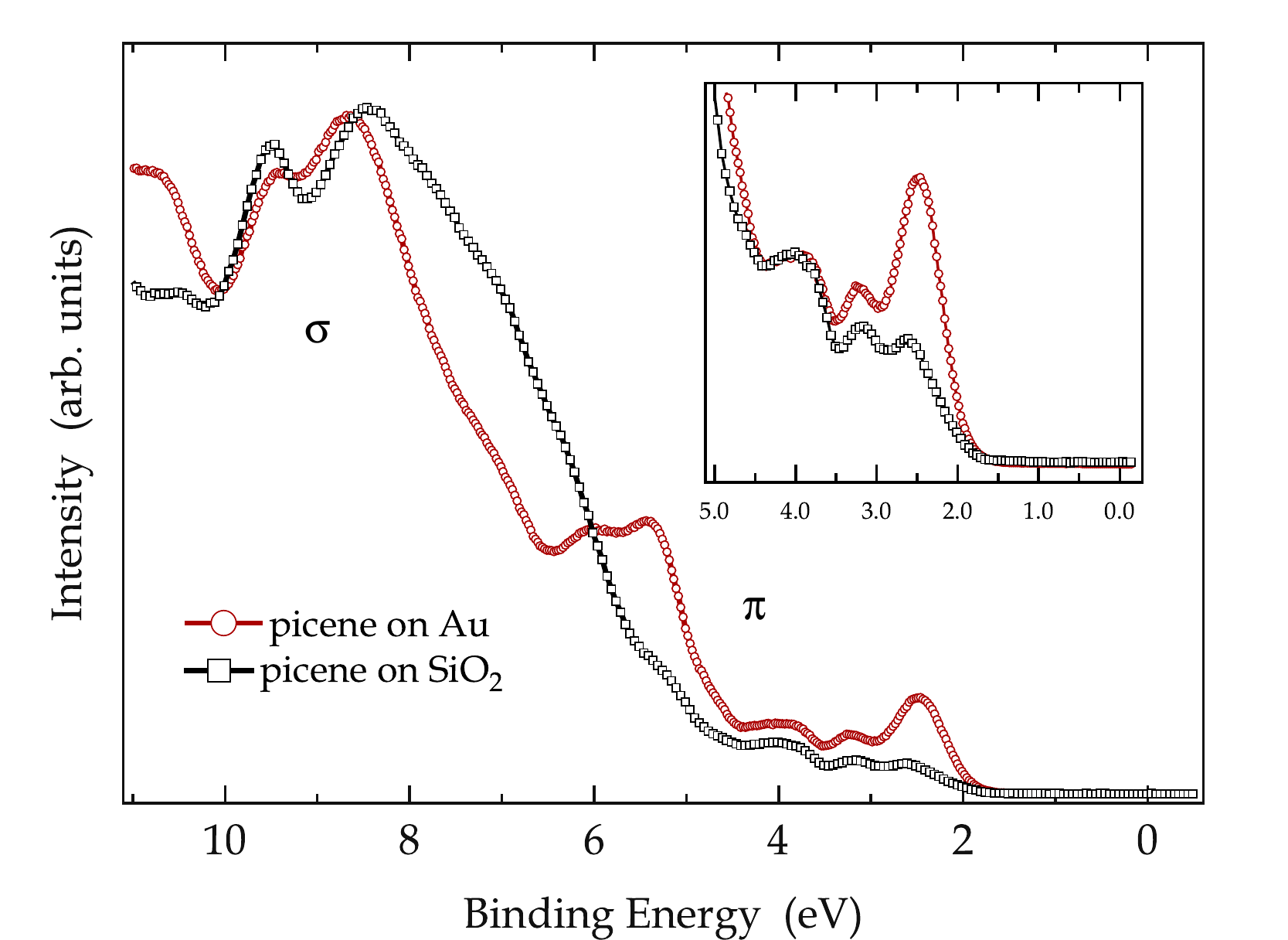} \caption{Valence band photoemission spectra of picene grown on gold and oxidized silicon, respectively. The
energy regions predominantly representing emission from $\sigma$ and $\pi$ states are labelled accordingly. The inset shows the energy region near
the Fermi level.} \label{f2}
\end{figure}

Thin films of picene and coronene (SIGMA-ALDRICH) for these measurements (about 5 nm thick) have been prepared by \emph{in situ} thermal evaporation onto a n-type Si
wafer kept at room temperature with a native oxide layer on top. Prior to organic film deposition the Si wafer was heated in the preparation
chamber for 60 min at 300 °C to remove surface contaminations. XPS was used to check the cleanliness of the heated substrate. A quartz
microbalance was used to monitor the thickness of the films which where grown with  typical deposition rates in the order of 4\,Å/min. It is
known that pentacene and other organic molecules show a tendency to lie flat on metal surfaces (e.g. gold)
\cite{Kang2003,Walzer1998,Eremtchenko2005,McDonald2006}, while they grow in a standing up manner on surfaces where the interaction with the substrate is
rather weak (e.\,g. oxidized silicon) \cite{Schwieger2004,Hatch2010,Hatch2009,Ruiz2003,Dimitrakopoulos1996,Thayer2005,Zheng2007}. Thus, the
picene and coronene films as grown in this study are composed of molecules with the long axis (picene) or their plane (coronene) standing
upright on the substrate. This is confirmed for picene in Fig.\,\ref{f2}, where we compare the valence band photoemission signal of picene
grown on gold and oxidized silicon in a normal emission geometry, respectively. These two spectra significantly differ in the relative
intensities in the energy regions around 9\,eV and 3-4\,eV, which predominantly represent emission from $\sigma$ and $\pi$ molecular states,
respectively. Such an intensity behavior is expected for mainly lying/standing molecules on the corresponding substrates due to the different
orientation of the responsible orbitals. \cite{Yannoulis1987,Hasegawa1996}

\par

Potassium was intercalated  in several steps by evaporation from commercial SAES (SAES GETTERS S.p.A.,Italy) getter sources at a pressure lower
than $6\times10^{-9}$\,mbar. The current through the SAES getter source was 5.8\,A and the distance to the sample was about 60\,mm. During
potassium addition, the sample was also kept at room temperature. The potassium concentration in the Picene and Coronene films was derived from
a comparison of the relative intensities of the K $2p$ and the C $1s$ core-level intensities which where corrected using the different
photoionization cross sections for carbon and potassium. As subshell photoionization cross sections we used 0.053 for K $2p$ and 0.013 for C $1s$
\cite{Yeh1985}. Taking into account deviation from ideal doping conditions and the uncertainty of these factors one arrives at an error
of the intercalation level of about $\pm 0.15$.

\section{Results and discussion}

\begin{figure}[h]
\centering
\includegraphics[width=0.49\linewidth]{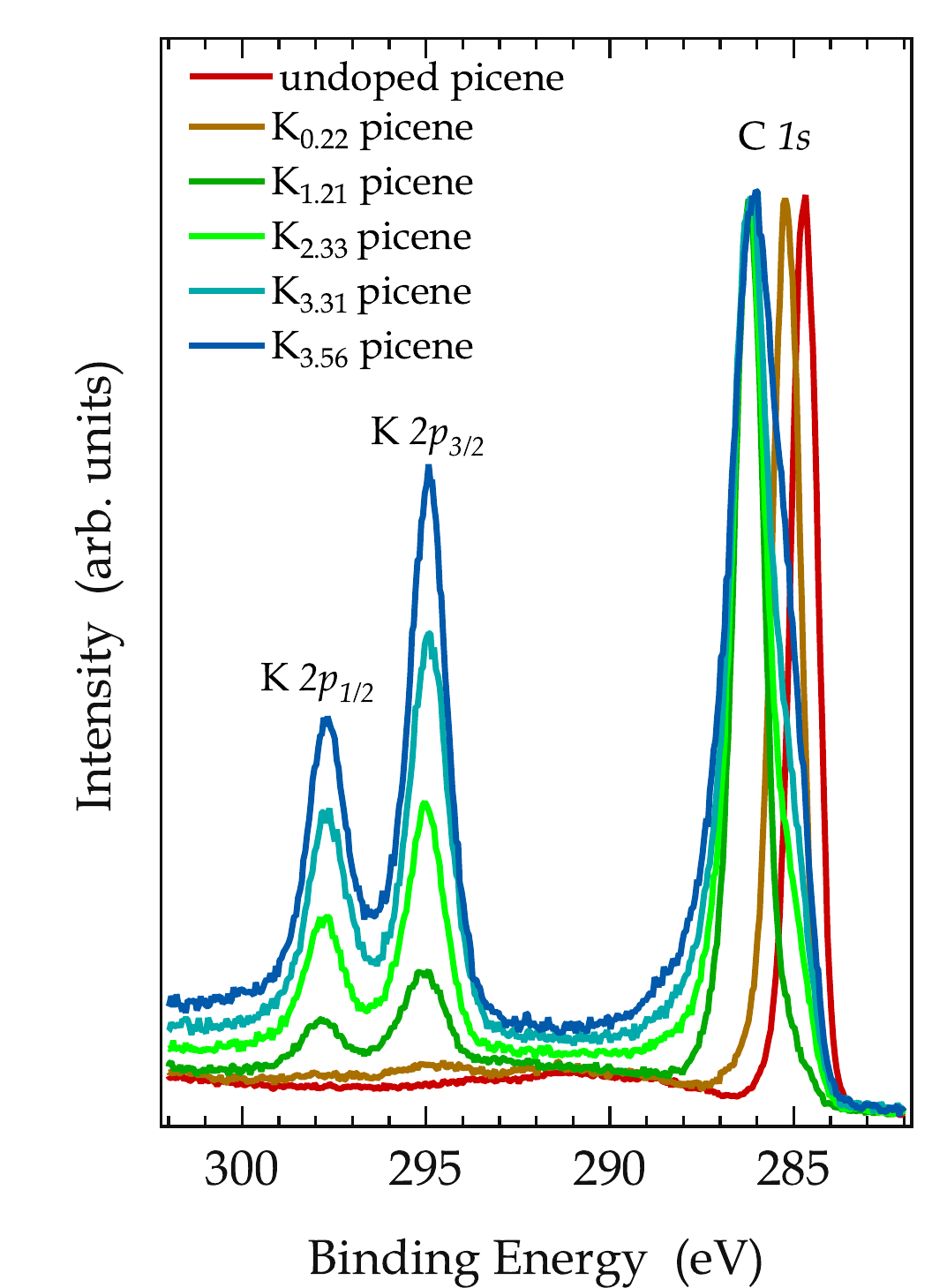}
\includegraphics[width=0.49\linewidth]{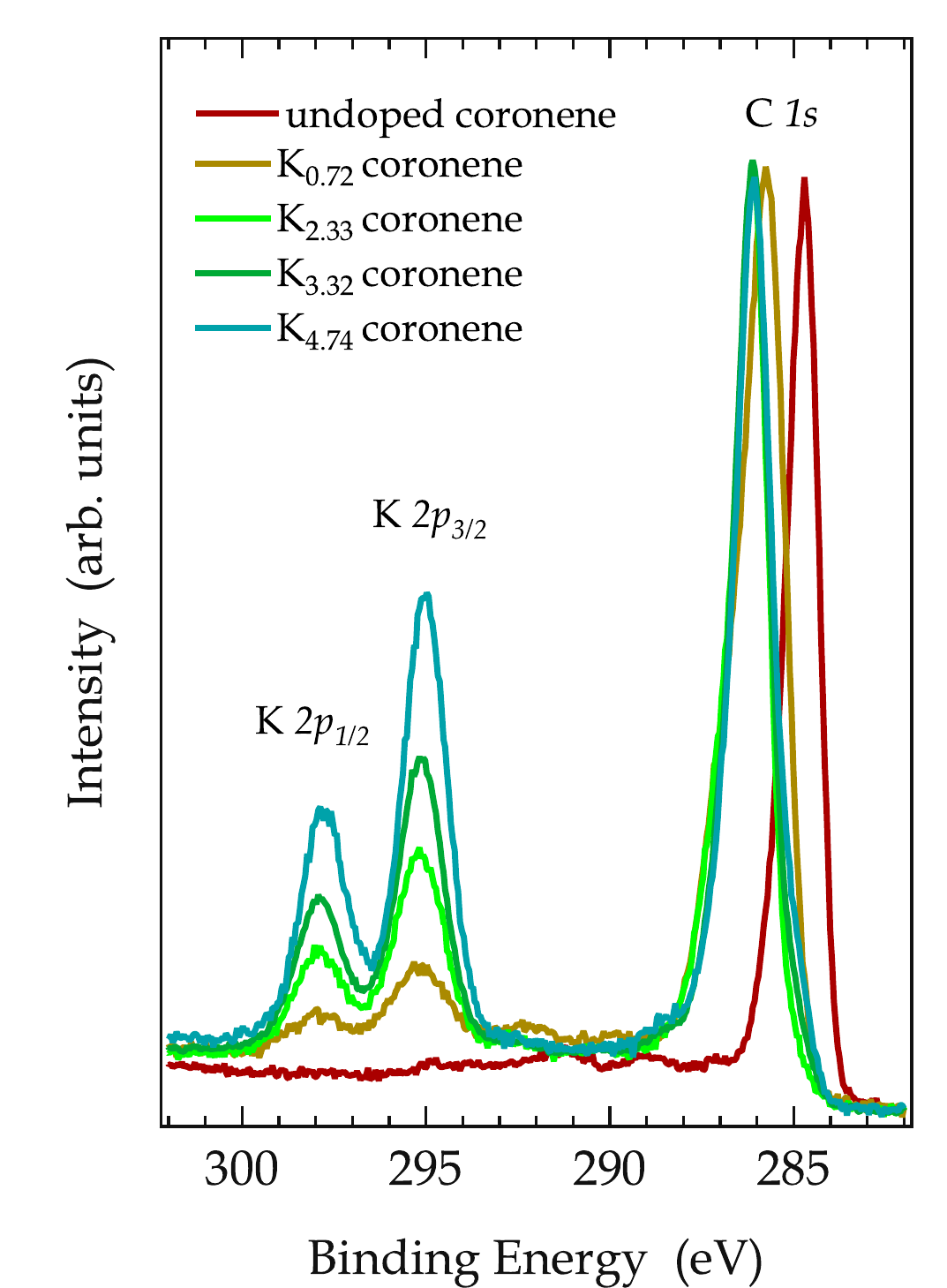}
\caption{C\,$1s$ and K\,$2p$  core level photoemission
spectra of picene (left panel) and coronene (right panel) as a function of potassium content $x$.} \label{f3}
\end{figure}

In Fig. \ref{f3} we show the evolution of the C 1$s$ and K 2$p$ core level photoemission data of the picene and coronene films as a function of
potassium content. For both undoped films the C 1$s$ core level spectrum consists of a single feature with a binding energy of 284.8\,eV (picene)
and 284.8\,eV (coronene) in agreement with previous studies \cite{Schroeder2002,Casu2007}. The relatively small peak width (about 0.85\,eV for
picene and 0.95\,eV for coronene) in both cases is evidence for the fact that the carbon atoms in both structures are rather equivalent
regarding electron densities and screening effects in the photoemission initial and final state. We note that for C$_{60}$, where all C atoms
are symmetrically equivalent, a line width of about 0.65\,eV was observed.\cite{POIRIER1991} At the high binding energy side, satellite features
can be seen which arise from excitations between $\pi$ and $\pi^*$ levels in the molecules due to screening of the final state core
hole.\cite{POIRIER1991,Golden1995,Benning1992,Enkvist1993}

\par

Initially, potassium addition results in a significant upshift of the spectra in binding energy by about 1.5\,eV for K$_x$picene and
K$_x$coronene, respectively. This upshift is directly related to the n-type doping process (i.\,e. the addition of electrons), which causes a
Fermi level shift towards the conduction band edge (the Fermi level represents zero binding energy). This upshift is equivalent to many other
studies of doped molecular films.\cite{Benning1992,Schwieger2001} Further, potassium doping is also seen by the appearance of spin-orbit split K 2$p$ core level structures
around 295\,eV and 297.8\,eV. The relative intensity of these potassium related features to that of the C 1$s$ structure is a direct measure of the
doping level (see above). After the initial energetic upshift the binding energy of the core level spectra is almost constant with further
doping, but the increasing K 2$p$ intensity clearly shows that the potassium content in the films grows until a doping level of about $x$ = 3.6 and
4.7, which represents doping saturation under the conditions applied in this work. Further potassium addition resulted in a K overlayer on top
of the doped molecular films.

\begin{figure}[h]
\centering
\includegraphics[width=.7\linewidth]{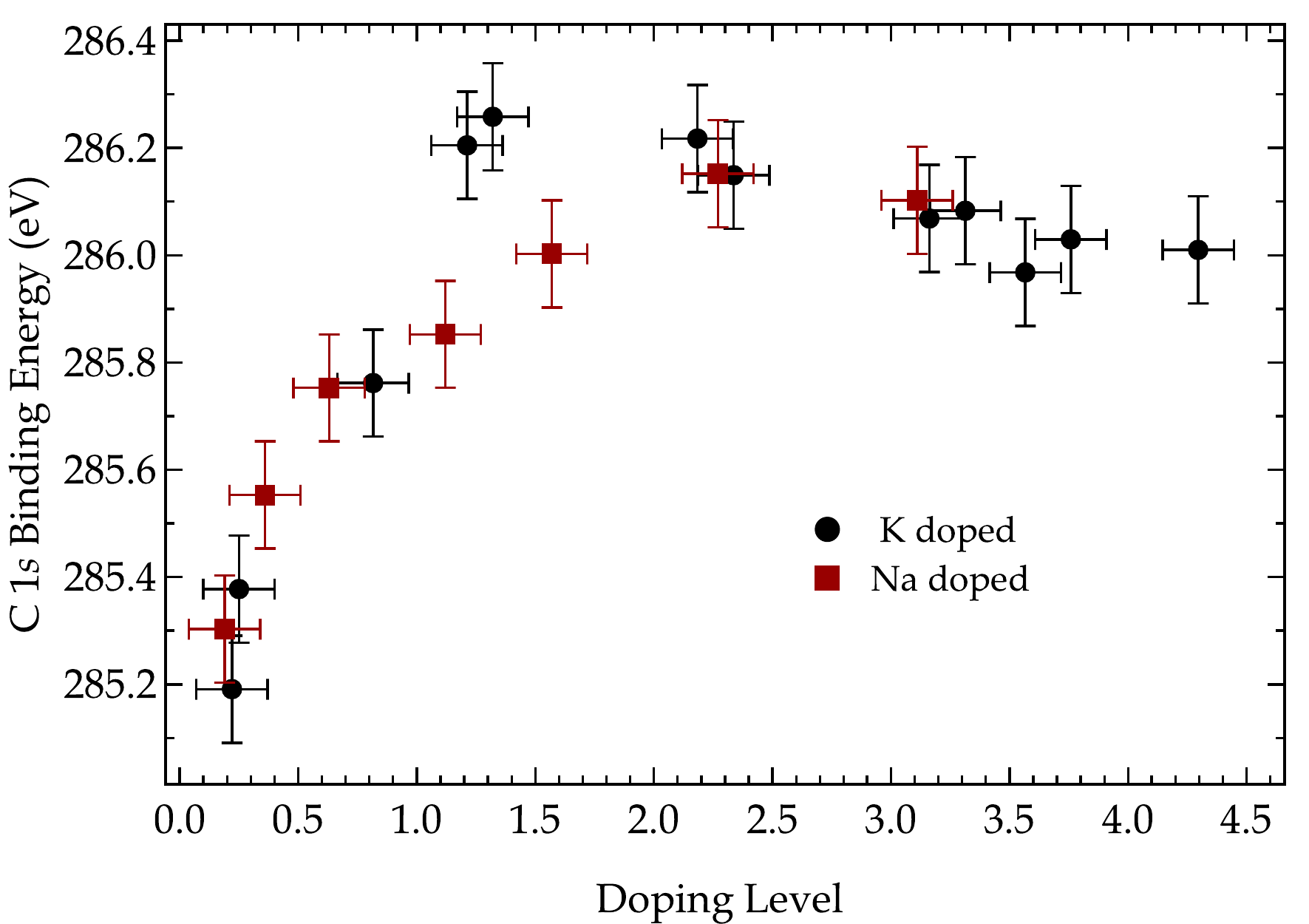}
\caption{Comparison of
the C 1$s$ binding energy of K$_x$picene and Na$_x$picene as a function of the doping level $x$.} \label{f4}
\end{figure}

 An equivalent behavior is observed for Na doping of picene as can bee seen in Fig.\ref{f4}, where we show a comparison of
the C 1$s$ binding energy of K$_x$picene and Na$_x$picene as a function of the doping level. This shows that in both cases the alkali metals
penetrate into the picene films and transfer an electron to the molecules.

\par

\begin{figure}[h]
\centering
\includegraphics[width=0.49\linewidth]{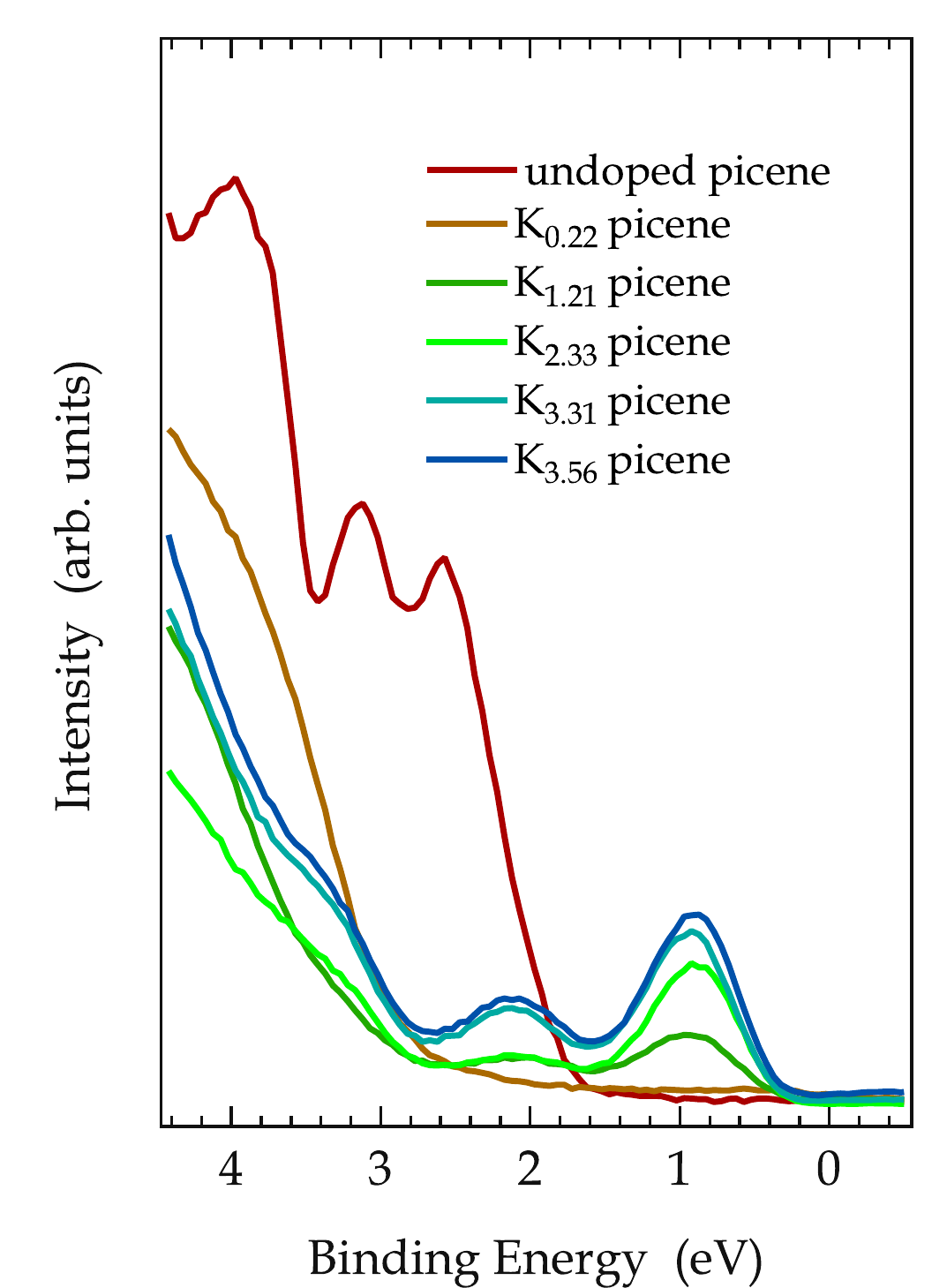}
\includegraphics[width=0.49\linewidth]{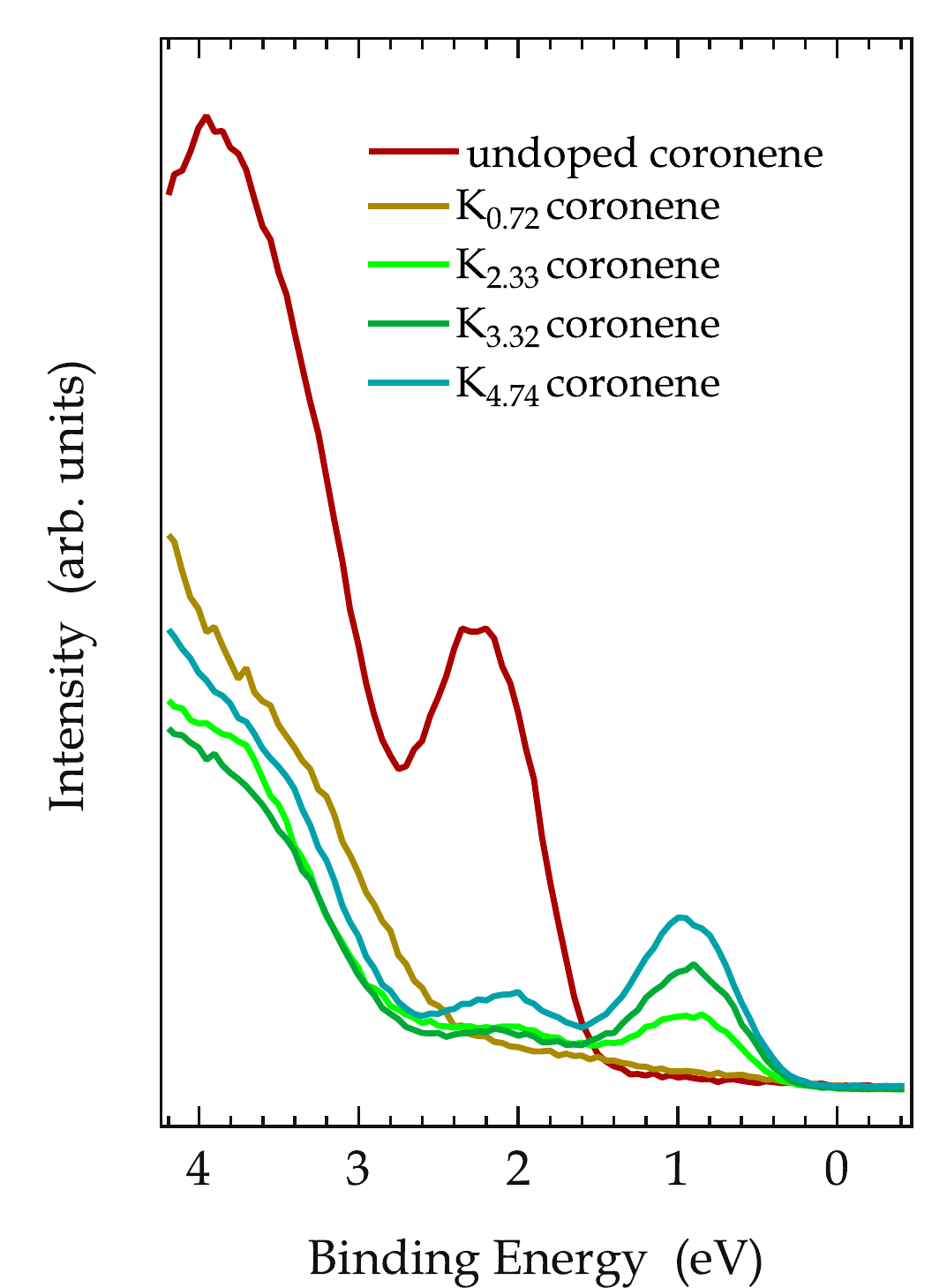}
\caption{Valanece band photoemission spectra of picene (left panel) and coronene  (right panel) as a function of potassium content $x$.} \label{f5}
\end{figure}

Fig. \ref{f5} presents the valence band data near the Fermi level and the changes induced by potassium doping for picene and coronene. The data
for the undoped materials are equivalent to those published previously.\cite{Ueno1978,Schroeder2002,Yamakado1998,Roth2010,Wang2011} Slight
doping again is observed as a strong upshift in the spectra, analogous to the core level data discussed above. Then, doping causes new spectral
structures close to the Fermi level, which arise from the filling of previously unoccupied states with the 4s electrons from potassium. For K
doped picene, we observe the appearance of in total three doping induced features at a binding energy of about 0.9, 2.1 and 3.3\,eV,
respectively. These energy positions are doping independent while the respective peak intensities grow with the doping level. The spectra for
doped coronene reveal two doping induced features at about 0.9 and 2\,eV, which also do not change in energy upon potassium addition but grow in
spectral weight.

\par

\begin{figure}[h]
\centering
\includegraphics[width=0.8\linewidth]{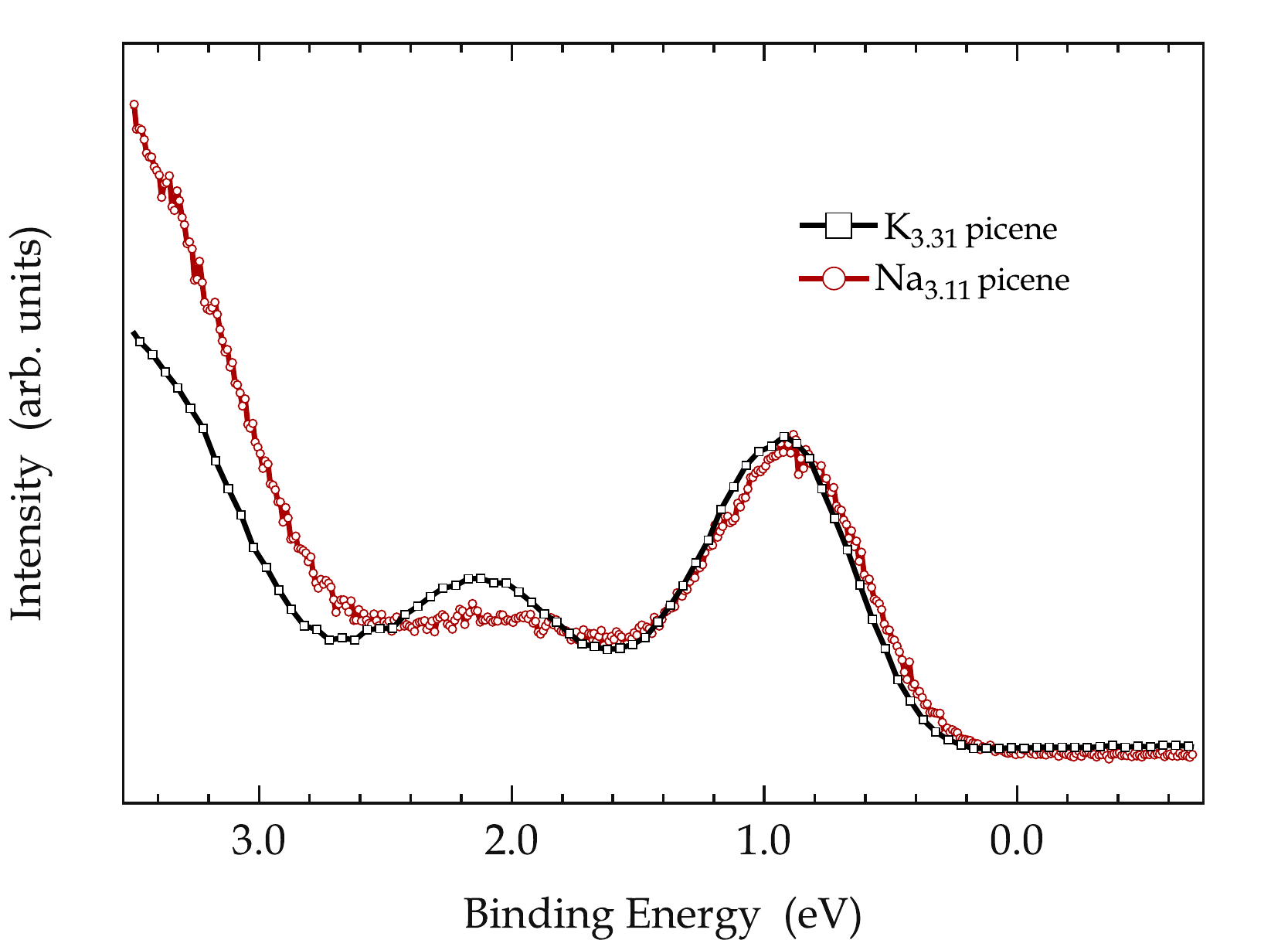}
\caption{Photoemission valence band spectra of K$_{3.3}$picene and Na$_{3.1}$picene on SiO$_2$ } \label{f6}
\end{figure}

Strikingly, doping of both molecular materials results in several new structures in the valence band for all doping levels. Moreover, at none of
the doping levels there is evidence for a finite intensity at the Fermi level which would represent a metallic doped picene or coronene film.
Further, we emphasize that very similar data are also observed for Na doped picene films as evidenced by Fig. \ref{f6}, which shows a comparison of the
valence band spectra of  K$_{3.3}$picene and Na$_{3.1}$picene. Apart from a small overall energy shift of 0.1\,eV, the data for the two alkali
metal doped films are characterized by similar spectral features around 0.95\,eV, 2.1\,eV and a shoulder centered at 3.2\,eV, but no intensity at the
Fermi level.

\par

We thus arrive at the surprising result that none of our films seems to become metallic upon K or Na doping, in strong contrast to the
observation of superconductivity in related crystals.\cite{Mitsuhashi2010} In the following, we discuss several scenarios that could be responsible for such a
discrepancy.

\par

In general, our photoemission studies are carried out on thin films that are grown \emph{in-situ} on a SiO$_2$ surface. Equivalent to other studies, we
expect that this results in a film structure where the molecules are growing with their long axis (picene) or their plane (coronene)
perpendicular to the substrate surface, and the molecular arrangement should then be close to that in bulk materials. We additionally note that
for undoped thin films of picene there is very good agreement between the measured data on such thin films and the density of states as obtained
from calculations of the bulk phase.\cite{Roth2010}

\par

The question however arises which doped phases are thermodynamically stable when picene or coronene are doped with potassium (or sodium). It is
well known from other molecular materials that particular doped phases might form while others are unstable. The most famous examples are
potassium doped C$_{60}$ compounds where for instance, stable K$_3$C$_{60}$, K$_4$C$_{60}$ and K$_6$C$_{60}$ phases have been found, but phases
with K$_2$C$_{60}$ or K$_5$C$_{60}$ do not exist.\cite{Poirier1995,Murphy1992,Rosseinsky1995} We thus cannot exclude that our film preparation
procedure does not result in films which represent the same crystal phase that has been observed to become superconducting. In this context it
is important to realize that the recent observation of superconductivity in alkali doped hydrocarbons required long time annealing procedure of
the material at about 440\,K in closed glass tubes \cite{Mitsuhashi2010}. Equivalent annealing cannot be applied in our ultra-high vacuum environment, the
application of temperatures above 420 K resulted in a loss of nearly the entire thin films.Moreover the existence of insulating crystal phases has also
been discussed recently on the basis of density functional theory calculations of potassium doped picene, and it has been predicted that phases
with K$_2$picene and K$_4$picene composition are band insulators, while K$_3$picene was found to be metallic.\cite{Kosugi2011a} As a consequence, the existance of alkali metal doped picene and coronene phases needs to be established experimentally in order to clarify their electronic ground state.

\par

Apart from crystal structure issues there is also structural changes of the molecular backbone itself upon charge addition. This is a natural
consequence of the filling of orbitals with anti-bonding character and the accompanied relaxation of the molecular structure, in other words the
presence of appreciable electron-phonon coupling with intra-molecular phonons. Indeed, such a coupling has been discussed to be responsible for
the formation of the superconducting ground state of picene.\cite{Subedi2011,Casula2011,Kato2011} On the other hand, for some molecular materials it has been reported that there is an energy gain connected to this structural relaxation upon charging which is strong enough in case of two charges on one molecule to
overcome the Coulomb repulsion, such that so-called bi-polarons are formed.\cite{Steinmuller1993,Ramsey1990,Koch2000,Murr1998,Logdlund1996} In
other words, upon charging or doping, molecules with two charges are the most stable species and singly charged molecules are not observed. This
scenario would be in good agreement with our data, since the observation of two (or more) valence band structures upon potassium addition is in
direct correspondence to the expectation in such a bi-polaron picture. \cite{Steinmuller1993,Ramsey1990,Koch2000,Murr1998,Logdlund1996} We note
that such bi-polarons might also be stabilized by the presence/attraction of the positively charged potassium (or sodium) ions in the crystal
lattice. However, while bi-polarons in principle could explain our data, they are in contrast to the observation of superconductivity, and
moreover, for structures such as K$_3$picene, which is reported to be a superconductor, it becomes rather unlikely that preferably picene$^{4-}$
molecules are formed and are energetically more favorable than picene$^{3-}$, since Coulomb repulsion on the individual molecules will get more
and more important with increasing charge.

\par

Finally, molecular crystals in general are materials with rather narrow energy bands, which is a direct result of the relatively weak
interaction between the individual molecules in the crystal. Moreover, the band width is in many cases comparable to the repulsion of two charge
carriers brought onto a molecule. Consequently, there is good evidence that molecular crystals can be regarded to be correlated
materials.\cite{Bruhwiler1993,Lof1992,Knupfer1994,Giovannetti2008,Tosatti2004,Knupfer2002} While superconductivity has been observed for
particular structures of K$_x$picene or K$_x$coronene, even a small structural difference could also be responsible for the different, metallic
or insulating, observed ground state at the same doping level, since this small difference might change the balance between the band width
(kinetic energy gain) and the Coulomb repulsion (energy cost for delocalization) in compounds with an integer doping level. A metal-insulator
transition is indeed known from the alkali-metal-intercalated C$_{60}$ materials, where a lattice expansion and symmetry lowering in previously
metallic K$_3$C$_{60}$ or a change of the lattice symmetry going from K$_3$C$_{60}$ to K$_4$C$_{60}$ results, in an insulating ground
state.\cite{Tou2000,Kitano2002,Durand2003,Knupfer1997} For aromatic hydrocarbons such as picene´and coronene crystals, it has also been
discussed that electronic correlation effects play an important role and that these crystals when doped with three potassium atoms per molecule
are close to a metal-insulator transition into a Mott insulating phase.\cite{Kim2011,Giovannetti2011,Nomura2011}

\section{Summary}

To summarize, we have investigated the electronic properties of potassium doped picene and coronene films using photoemission spectroscopy. Our
studies reveal a complex behavior upon doping. For both molecular materials we observe a clear shift of all photoemission signals at the
beginning of the doping process. This shift documents the upshift of the Fermi level towards the conduction band edge as negative charge
carriers (electrons) are introduced. Furthermore, the addition of charges leads to the appearance of three (picene) or two (coronene) additional
valance band structure in the former gap, respectively. These arise from molecular orbitals that now are filled as a consequence of the doping
process. Intriguingly, for none of the films we observe emission from the Fermi level, i.\,e. none of the films in our studies becomes metallic.
An equivalent behavior is also found for sodium doped picene films. In view of the possibility of several alkali metal doped phases in these 
hydrocarbon crystals, it is required to establish the phase diagram in order to be able to clarify the electronic ground state. Moreover, electronic correlations and 
electron-phonon coupling most likely are essential for as complete understanding of these materials and many relatives.

\begin{acknowledgments}
We thank A. Ruff, M. Sing and R. Claessen for fruitful discussions and R. Sch\"onfelder, R. H\"ubel and S. Leger for technical
assistance. This work has been supported by the Deutsche
Forschungsgemeinschaft (grant number KN393/14). 
\end{acknowledgments}


%

\end{document}